\documentstyle[sprocl]{article}

\def\sst{\scriptscriptstyle}

\def\etap{\eta^\prime}

\def\be{\begin{equation}}
\def\ee{\end{equation}}
\def\bea{\begin{eqnarray}}
\def\eea{\end{eqnarray}}

\begin{document}

\title{Large and Fun CP Violation in B Meson Decays: Where to Search?}

\author{George W.S. Hou
\footnote{
Permanent address: Dept. of Physics,
National Taiwan Univ., Taipei, Taiwan, R.O.C.}
}

\address{Brookhaven National Lab, Physics Department, Bldg. 510A\\ 
Upton, NY 11973, USA\\E-mail: wshou@quark.phy.bnl.gov}

\maketitle\abstracts{ 
The best place to search for direct CP violation is 
the already observed charmless $b\to s$ modes.
In SM with FSI, $a_{CP}$ in $K\pi$ modes could be as large as
20--30\% but differ in sign between $K^-\pi^+/K^-\pi^0$ and $\bar K^0 \pi^-$.
We illustrate possible New Physics effects
that could lead to $a_{CP}$ of order 40--60\%
in $K\pi$ and $\phi K$ modes distinguishable from FSI.
$b\to s\gamma$ modes can also exhibit interesting asymmetries.
}

\section{Motivation}
1997 can be called the Year of the Strong Penguin: 
a handful of two-body charmless B decays were observed~\cite{TwoBody} 
by CLEO for the first time, giving firm indication
that strong penguins are dominant.
Something completely unexpected also emerged in $\etap$ modes:
Not only exclusive modes are very sizable,
semi-inclusive $B\to \etap + X_s$ with fast $\etap$
was found~\cite{etapXs} to be close to $10^{-3}$.

We concern ourselves here with 
direct CP violation in these modes.
Given the statistics, 
the $a_{CP}$ reach is only $\sim 100\%$ at present.
But, as B Factories are turning on soon, 
in 2--5 years this could go down to $30\%$ to 10\%.
The modes that are already observed so far would 
certainly be the most sensitive probes.
What physics do they and can they probe?

What has been observed so far are charmless $b\to s$ decays.
The ordering $K\pi > \pi\pi$
clearly indicates that strong penguin $>$ tree.
We recall that in SM,
$a_{CP}$ for pure penguin $b\to s$ transitions are suppressed
by the factor 
\be
{\rm Im}\,(V_{us}V_{ub}^*)/\vert V_{cs}V_{cb}^*\vert
  \simeq\eta\lambda^2 < 1.7\%,
\ee
so $a_{CP} > 10\%$ in pure penguin modes 
would imply New Physics!
We therefore have a discovery window in the next few years for 
beyond SM (BSM) effects.
The question then is: What BSM?
{\it Is large $a_{CP}$ possible in $b\to s$ modes?}
Rather than trying to be exhaustive, 
we wish to demonstrate that CP asymmetries in $b\to s$ transitions
can indeed be large with simple extensions of SM, 
and sometimes even within SM.

New Physics will be illustrated with {\it large color dipole} 
$bsg$ coupling
\begin{eqnarray}
-{\frac{G_{F}}{\sqrt{2}}}{\frac{g}{16\pi^{2}}}V_{tb}V_{ts}^{*} \, 
c_{8}\, \bar{s}\sigma _{\mu \nu }G^{\mu \nu} m_{b}\,(1+\gamma_{5})b.
\end{eqnarray}
In SM one finds $c_8^{\sst\rm SM}(m_b) \simeq -0.3$
which leads to $b\to sg$ (with $g$ on-shell) $\sim 0.2\%$,
a small rate that is usually neglected.
But because $b\to sg$ just does not give tangible signatures, 
our experimental knowledge of the strength of $c_8$ is actually rather poor.
In fact, the long-standing ``deficit" in 
$B$ decay semileptonic branching ratio (${\cal B}_{s.l.}$) 
and charm counting ($n_{C}$) point towards the possibility of 
sizable $b\to sg$ in Nature.~\cite{bsg,kagan}
If $b\to sg \approx 10\%$, which implies $c_8 \sim 2$,
${\cal B}_{s.l.}$ and $n_C$ can each be lowered by that amount and the 
problems would go away.
A recent CLEO bound~\cite{bsglimit} gives $b\to sg < 7\%$ at 90\% C.L.,
which comes indirectly from the study of $B\to D\bar D K + X$ decay.
But even if one takes this seriously there is still much room, 
and $b\to sg$ at 1--5\% would be very hard to rule out.
What we stress here is:
if $c_8$ is large in Nature, it must be coming from New Physics
and should carry naturally a KM-independent CP violating phase.

The idea of an enhanced $bsg$ color dipole coupling and its associated 
new physics CP phase has been applied to $B\to \eta^\prime + X_s$ decay.
We have advocated that 
the $g^*g\eta^\prime$ anomaly coupling mechanism~\cite{AS,HT} is needed 
to account for the energetic $\eta^\prime$ (or equivalently, 
the recoil $m_{X_s}$) spectrum.
Then, with new CP phase $\sigma$ in $c_8 \cong 2e^{i\sigma}$ 
interfering with absorptive parts from usual $c_{3-6}$ penguin coefficients,
$a_{CP}$ in inclusive $m_{X_s}$ spectrum could be at 10\% level.~\cite{HT}

We explore here \cite{HHY} the general impact of a large color dipole coupling 
on CP asymmetries in charmless $b\to s$ decays.
If $b\to sg$ rate is really of order 1--10\% in Nature, 
even if this rate itself is hard to measure,
other charmless $b\to s$ decays must be 
affected through interference effects.

\section{Model of Unconstrained CP Phase}

To have large $b\to sg$ and evade $b\to s\gamma$ constraint at the same 
time, one needs additional source for radiating gluons but not photons.
Gluinos ($\tilde g$) fit the bill nicely.
In SUSY one usually sets soft squark mass terms to be ``universal"
to suppress FCNC and to reduce the number of parameters.
But it has been shown~\cite{kagan,susy} that 
nonuniversal soft $m_{\tilde d_j}$ masses could give large 
$c_8$ without violating the $b\to s\gamma$ constraint.
In previous studies, however, the possibility of new CP phases were not 
considered.

As an existence proof, let us consider a 
minimal model of $\tilde s - \tilde b$ mixing, 
the simplest would be $LL$ mixing~\cite{HH} which mimics SM couplings, 
but one could also have $RR$ or $LR$ mixing models.~\cite{CHH} 
The phase of $d_i$ quarks are fixed by gauge interaction, and there is 
just one mixing angle $\theta$ and one phase $\phi$.
Since this mixing involves only the second and third generations,
one evades low energy bounds that involve first generation quarks, such 
as neutron edm, the $K$ system, and even $B_d$-$\bar B_d$ mixing.
This is a natural model that is tailor made for generating large effects 
in $b\to s$ transitions (as well as $B_s$ mixing)!

\section{Direct CP Violation in Inclusive $b\to s\bar qq$ Decay}

The theory of inclusive decays are cleaner since one can use the 
quark/parton language. The absorptive parts arrising from short distance 
perturbative rescattering~\cite{BSS} can be used and one is insensitive 
to long distance phases. However, experimental detection poses a 
challenge, unless partial reconstruction techniques can be made to work.

Since penguins dominate charmless $b\to s$ decays, 
one is interested in CP violation in pure penguin processes 
such as $b\to s\bar dd$ and $s\bar ss$. 
But since these rates and 
asymmetries occur at ${\cal O}(\alpha_S^2)$, 
care~\cite{GH} has to be taken in treating CP violation in 
the $b\to s\bar uu$ mode, 
which has the distinction of receiving also the tree contribution. 
Although the tree amplitude alone does not lead to CP violation, 
while tree--penguin interference occurs only at ${\cal O}(\alpha_S)$,
to be consistent with treating pure penguin CP asymmetries, 
one needs to take into account the absorptive part carried by 
the gluon propagator (bubble graph) associated with the penguin. 
This ${\cal O}(\alpha_S^2)$ tree--penguin interference term is needed to 
maintain CPT and unitarity in rates and hence $a_{CP}$.

The above discussion has been stated in terms of ``full" theory (exact 
loop calculation) to lowest relevant order in $\alpha_S$. Since QCD 
corrections are important and relatively well developed by now, we adopt 
an operator language in computing inclusive rates. We start from the 
effective Hamiltonian
\begin{eqnarray}
H_{\rm eff} &=& {4G_F\over \sqrt{2}} \left[
V_{ub}V_{us}^*(c_1O_1 + c_2 O_2)
-V_{ib}V_{is}^* \, c^i_j O_j\right],
\label{Heff}
\end{eqnarray}
with $i$ summed over $u,c,t$ and $j$ over $3$ to $8$.
The operators are defined as
\begin{eqnarray}
&& O_1 = \bar u_\alpha \gamma_\mu L b_\beta \,
        \bar s_\beta \gamma^\mu L u_\alpha,
\ \ \;
O_2 = \bar u \gamma_\mu L b \, \bar s \gamma^\mu L u,
\nonumber\\
&& O_{3,5} = \bar s \gamma_\mu L b \, \bar q \gamma^\mu {L(R)} q,
\ \ O_{4,6} = \bar s_\alpha\gamma_\mu L b_\beta \,
      \bar q_\beta \gamma^\mu {L(R)} q_\alpha,
\nonumber\\
&& \tilde O_8 = {\alpha_s\over 4\pi}\, \bar s i\sigma_{\mu\nu} T^a
                {m_b q^\nu\over q^2} Rb\, \bar q \gamma^\mu T^a q,
\end{eqnarray}
where $\tilde O_8$ arises from the dimension 5 color dipole $O_8$ 
operator of Eq. (2), and $q= p_b-p_s$.
We have neglected electroweak penguins for simplicity.
The Wilson coefficients $c_j^i$ are evaluated to NLO order
in regularization scheme independent way,
for $m_t = 174$ GeV, $\alpha_s (m_Z^2) = 0.118$ and $\mu = m_b = 5$ GeV.
Numerically,~\cite{desh-he} $c_{1,2} = -0.313,\ 1.150$, 
$c_{3,4,5,6}^t = 0.017,\ -0.037,\ 0.010,\ -0.045$,
and $c_8^{\sst\rm SM} = c_8^t-c_8^c= -0.299$.
We note that 
$c_{1,2}$ are resummations of series starting at ${\cal O}(\alpha_S^0)$, 
while $c_{3-6}$ start at ${\cal O}(\alpha_S^1)$
which is reflected in their relative smallness. 
However, 
one power of $\alpha_S$ is factored out by convention in defining $O_8$, 
hence $c_8$ starts at ${\cal O}(\alpha_S^0)$ and its size is 
comparable to $c_1$ within SM.
One has to keep track of the {\it relevant leading order} in $\alpha_S$
when comparing with ``full theory" approach discussed earlier.

To get absorptive parts, we add
$c^{u,c}_{4,6}(q^2) = -Nc_{3,5}^{u,c}(q^2)=-P^{u,c}(q^2)$
for $u$, $c$ quarks in the loop, where
\[
P^{u,c}(q^2) =
 {\alpha_s \over 8\pi} c_2 \left({10\over 9} + G(m_{u,c}^2,q^2)\right),
\]
and
\[
G(m^2,q^2) =
4\int x(1-x) dx\, {\rm ln}\, {m^2 - x(1-x)q^2\over \mu^2}.
\]
To respect CPT/unitarity at ${\cal O}(\alpha_S^2)$,
for $c^t_{3-8}$ at $\mu^2 = q^2 < m_b^2$, we substitute
\begin{eqnarray}
{\rm Im}\, c_8 &=& {\alpha_s\over 4\pi}  c_8 \,
                      \sum_{u,d,s,c} {\rm Im}\, G(m_i^2,q^2), 
\nonumber\\
{\rm Im}\, c^t_{4,6} &=& -N{{\rm Im}\, c^t_{3,5}}
={\alpha_s\over 8\pi} \left[c^t_3 {\rm Im}\, G(m_s^2,q^2)
+ (c^t_4+c^t_6)
    \sum_{u,d,s,c}{\rm Im}\, G(m_i^2,q^2)\right],
\nonumber
\end{eqnarray}
when interfering with the tree amplitude.
We note that the use of operator language can be misleading at this stage 
since the absorptive parts are not resummed while the Wilson coefficients are. 
One could easily lose track of $\alpha_S$ counting that is 
needed for maintaining CPT/unitarity 
if one thinks too naively in effective theory language.

Having made all these precautions, we can square amplitudes 
in a straightforward manner to obtain rates and arrive at the asymmetries.
Since at lower order one has $b\to sg$ decay,
the $\vert c_8\vert^2$ term has a $\log q^2$ pole behavior.
We regulate it by simply cutting it off at 1 GeV.
Fig. 1(a) gives the rates for 
$b\to s\bar dd$ (solid) and $\bar b\to \bar sd\bar d$ (dashed)
vs. y$ = q^2/m_b^2$.
The SM result does not show a prominent low $q^2$ tail
since $b\to sg$ is small,
and the asymmetry comes mostly from below $c\bar c$ threshold.
For larger $q^2$ the $a_{CP}$ is GIM suppressed.~\cite{GH}
The SM asymmetry is indeed tiny.
For new physics enhanced $c_8 = 2e^{i\sigma}$,
we consider the cases for $\sigma = \pi/4$, $\pi/2$ and $3\pi/4$.
Besides a very prominent low $q^2$ tail since $b\to sg$ is now $\sim 10\%$, 
the salient feature is the 
rather large rate asymmetries above $c\bar c$ threshold.
The reason is because the new physics $\sigma$ phase now evades the 
SM constraint of Eq. (1), and the $c_8$ amplitude interferes with 
standard $c_{3-6}$ penguins which carry the absorptive parts 
due to (perturbative) $c\bar c$ rescattering, 
but the $u\bar u$ rescattering absorptive part is suppressed by $V_{ub}$.
Note that for $\sigma = \pi/4,\ 3\pi/4$ one has constructive, 
destructive interference, respectively. 
For the latter case, the overall rate is close to SM but the asymmetries 
are much larger, reaching 30\% for large $q^2$!

\begin{figure}[t]
\includegraphics{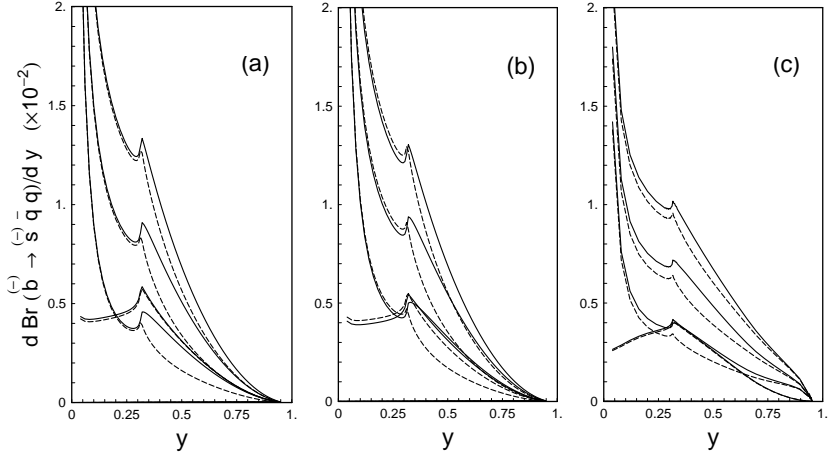}
\vskip 5.8cm
\caption{
Inclusive rate vs. y$ = q^2/m_b^2$ for
(a) $b\to s\bar dd$, (b) $s\bar uu$ and (c) $s\bar ss$ decays
(solid) and $\bar b$ decays (dashed). Curves with
prominent small $y$ tail are for $c_8 = 2e^{i\sigma}$  
with $\sigma = \pi/4$ (top), $\pi/2$ (middle), $3\pi/4$ (bottom), 
while the other is SM result.
}
\end{figure}
%
%

For $b\to s\bar uu$ the tree diagram also contributes, and 
one has to include the absorptive part in gluon propagator
as discussed earlier.
Because of this, the rate asymmetries in SM occur 
both below and above $c\bar c$ threshold, as can be seen in Fig. 1(b). 
Each are larger than the $b\to s\bar dd$ case
but are of opposite sign hence they tend to cancel each other.~\cite{GH}
If $c_8$ is enhanced, however, the dominant mechanism is 
again interference between $c_8$ and the usual penguins, 
hence the results are similar to the $b\to s\bar dd$ case.
For $b\to s\bar ss$ mode, 
one has to take into account identical particle effects.
As seen in Fig. 1(c), 
this leads to the peculiar shapes at large and small $q^2$, 
and the asymmetry is now smeared over all $q^2$. 
But otherwise it is similar to the $b\to s\bar dd$ case.

The integrated inclusive results are summarized in Table 1.

\begin{table}[htb]
\caption{Inclusive $Br$ (in $10^{-3}$)/$a_{CP}$ (in \%) for SM and for 
$c_8 = 2 e^{i\sigma}$.}
\vspace{0.15cm}
\begin{center}
\footnotesize
\begin{tabular}{|c|c|c|c|c|c|c|c|}
\hline
&SM &$\sigma$ =\ \ \ \ \ \  0 \ \ \ \ \ & ${i\pi/4}$ & 
${i\pi/2}$ & ${i3\pi/4}$ &$i\pi$  \\
\hline $b\to s \bar d d$ & 2.6/0.8   & \ \ \ \ \ \ 8.5/0.4 & 7.6/3.4 &
5.2/6.5 & 2.9/8.1 & 1.9/0.5 \\
\hline $b\to s \bar u u$ & 2.4/1.4   & \ \ \ \ \ \ \ 8.1/-0.2 & 7.5/2.6 
&
5.5/5.6 & 3.2/8.1 & 2.0/3.5 \\
\hline $b\to s \bar s s$ & 2.0/0.9 & \ \ \ \ \ \ 6.9/0.4 & 6.2/3.2 &
4.4/6.0 & 2.6/7.1 & 1.8/0.4 \\ 
\hline
\end{tabular}
\end{center}
\end{table}

\section{Direct CP Violation in Exclusive Charmless Hadronic Modes}

The exclusive modes are more accessible experimentally, 
as evidenced by the handful of observed modes.
Unfortunately, the theory is not clean.
One has to evaluate all possible hadronic matrix elements of 
operators in Eq. (3).
Faced with CLEO data, it has become popular~\cite{Neff} to use 
$N_{\sst\rm eff}$ rather than the value of 3 as dictated by QCD.
Although it is a measure of deviation from factorization,
it becomes in reality a new process dependent fit parameter.
One is still subject to usual approximations and inprecise knowledge of 
form factors and the $q^2$ value to take.
CP asymmetries are especially sensitive to the latter.
At the rate level, the $K\pi$ modes are approximately manageable,
but the $\eta^\prime K$ and $\omega K$ modes seem high 
while the $\phi K$ mode seems low.
Thus, even introducing $N_{\sst\rm eff}$ as a new parameter,
there are problems everywhere already in rate.
A new development~\cite{cleo1} in 1998 is that the
$B^- \to K^-\pi^0$ mode has been observed, while the
$B^- \to \bar K^0 \pi^-$ rate came down considerably.
One has now three measured $K\pi$ modes and their rates
are all around $1.4\times 10^{-5}$.

We shall not discuss the $\eta^\prime$ modes here since 
it must have a large contribution from anomaly mechanism 
and is rather difficult to treat.
But we do wish to explore whether an enhanced $c_8$ could improve 
agreement with experiment.
Before we do so, however, we point out that the $K\pi$ modes offer 
a rather interesting subtlety: 
they in general have two isospin components and 
exhibit larger $a_{CP}$ within SM, 
and they are very sensitive to final state interaction (FSI) phases.
As shown in Ref. [12] but now put in terms of the angle $\gamma$, 
in the absence of FSI phases one finds for $B^- \to K^-\pi^0$ mode
\be
a_{uu} \propto {\#_1\sin\gamma\over \vert\#_2 - \#_3\cos\gamma\vert^2},
\label{auu}
\ee
where $\#_1$ comes from interference, while 
$\#_2$ and $\#_3$ come from penguin and tree $b\to s\bar uu$ amplitudes, 
respectively. $\#_3$ and the dispersive part of $\#_2$ have the same sign.
At the time of Ref. [12], $\cos\gamma < 0$ was favored,
while $\sin\gamma$ was smaller than today,
hence $a_{uu}$ was not very large.
The present~\cite{stocchi} preferred value is $\gamma \sim 64^\circ$,
however, and one now has destructive rather than constructive interference.
Hence, one not only gains from $\sin\gamma\sim 0.9$ in the numerator,
there is also extra enhancement from the denomenator of Eq. (\ref{auu}),
and $a_{uu}$ as large as 10\% is possible.
Furthermore, since one can write 
$(\bar su)(\bar ub) = [(\bar su)(\bar ub) + (\bar sd)(\bar db)]/2
                     +[(\bar su)(\bar ub) - (\bar sd)(\bar db)]/2$,
there is in general two isospin components
from the tree level $O_1$ and $O_2$ operators.
These two isospin amplitudes may  develope soft FSI phases
that are different from each other.
If such is the case, it could overrun the perturbatively generated 
(hence $\alpha_S$ suppressed) absorptive phases,
and much larger CP asymmetries can be achieved.
While this is good news for CP violation search in general, 
it is bad news for search of new physics.
Can one distiniguish between new physics effects and SM with large FSI 
phases? The answer is yes, if one compares several modes.

Let us give a little more detail for sake of illustration.
We separate the $\bar B^0 \to K^-\pi^+$ amplitue into two isospin 
components, $A = A_{1/2} +  A_{3/2}$.
Since color allowed amplitudes dominate,
$N_{\sst\rm eff} \simeq N = 3$.
Defining $v_i = V_{is}^*V_{ib}$ and assuming factorization, we find,
\begin{eqnarray}
&& A_{1/2} =
i{G_F\over \sqrt{2}} f_K F_0\; (m_B^2-m_\pi^2)
\left\{v_u \left[ {2\over 3} \left({c_1\over N} + c_2\right)
-{r\over 3}\left(c_1+{c_2\over N}\right) \right] \phantom{c_5^j\over N} \right .
\nonumber\\
&& - v_j  \left. \left[ {c_3^j\over N} + c_4^j
+{2m_K^2\over (m_b-m_u)(m_s+m_u)}\left({c_5^j\over N} +c_6^j\right)\right]
-v_t {\alpha_s\over 4\pi} {m_b^2\over q^2} c_8 \tilde S_{\pi 
K} \right\},~
\label{kpi1}
\end{eqnarray}
where $F_0 = F_0^{B\pi}(m^2_K)$ is a BSW form factor,
$\tilde S_{\pi K} \sim -0.76$ is a complicated form factor 
normalized to $F_0$,
and $r$ is some ratio of $B\to K$ and $B\to \pi$ form factors
and a measure of SU(3) breaking.
For $A_{3/2}$ one sets $c^j_{3-8}$ to zero
and substitutes $2/3,\ -r/3 \longrightarrow 1/3,\ r/3$.
The $K^- \pi^0$ mode is analogous,
with modifications in  $A_{3/2}$ and an overall factor of $1/\sqrt{2}$.
The penguins contribute only to $A_{1/2}$,
hence the naively pure penguin $B^- \to \bar K^0 \pi^-$ amplitude
has just Eq. (\ref{kpi1}) with $c_{1,2}$ set to zero.
Note that the $c_{5,6}$ effects are sensitive to
current quark masses
because of effective density-density interaction.
The absorptive parts for $c^j_{3-8}$ are evaluated at
$q^2 \approx m_b^2/2$ which favors large $a_{CP}$, but
$q^2$ could be~\cite{GH} as low as $m_b^2/4$.
We plot in Fig. 2(a) and (b) 
the branching ratio ($Br$) and $a_{CP}$ vs. angle $\gamma$.
For $K^- \pi^{+,0}$, $a_{CP}$ peaks at the sizable $\sim 10\%$
just at the currently favored \cite{stocchi} value of
$\gamma \simeq 64^\circ$.
But for $\bar K^0 \pi^-$, $a_{CP} \sim \eta\lambda^2$ is very small.
We have used $m_s(\mu = m_b) \simeq 120$ MeV
since it enhances the rates.
With $m_s(\mu = 1$ GeV) $\simeq$ 200 MeV,
the rates would be a factor of 2 smaller.
We find $K^- \pi^+$, $\bar K^0 \pi^-$, $K^- \pi^0
 \sim 1.4,\ 1.6,\ 0.7 \times 10^{-5}$, respectively.

To illustrate the effect of FSI, 
we now write $A = A_{1/2} +  A_{3/2} e^{i\delta}$
and plot in Fig. 2(c) and (d) the $Br$ and $a_{CP}$ vs. $\delta$ for
$\gamma = 64^\circ$.
The rate is not very sensitive to $\delta$ which reflects
penguin dominance, but
$a_{CP}$ can now reach $20\%$, even 30\% for $K^-\pi^0$.
We stress that the naively pure penguin $\bar K^0 \pi^-$ mode
is in fact also quite susceptible to FSI phases
as it is the isospin partner of $K^-\pi^0$,
which definitely receives tree contributions.
The $B^- \to \bar K^0 \pi^-$ mode can receive tree contributions
through FSI rescattering.
Comparing Fig. 2(b) and (d), $a_{CP}$ in this mode can be
{\it much larger} than the naive factorization result.
However, the $a_{CP}$ for  $\bar K^0 \pi^-$ and $K^-\pi^+$ are out of 
phase,
hence, comparing the two cases
can give information on the FSI phase $\delta$.

\begin{figure}[htb]
\includegraphics{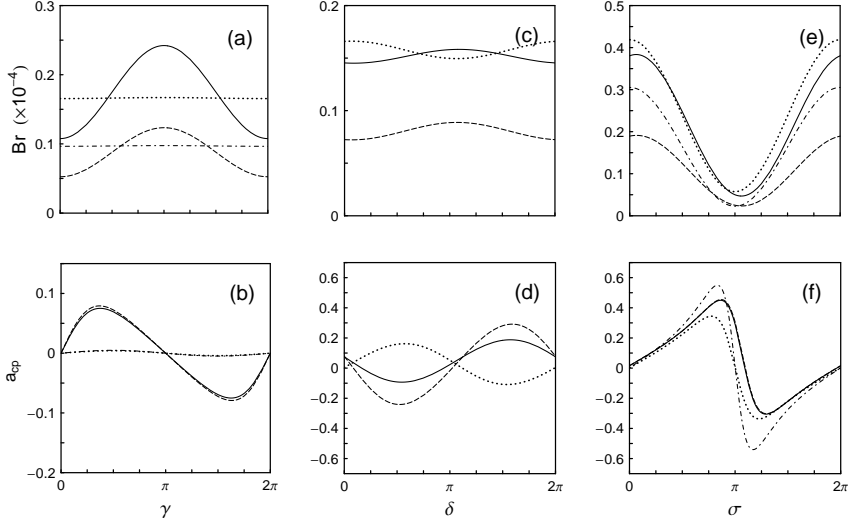}
\vskip 6.6cm
\caption { 
$Br$ and $a_{CP}$ vs. (a), (b) SM unitarity angle $\gamma$, 
(c), (d) FSI phase $\delta$ for $\gamma = 64^\circ$, and 
(e), (f) new physics phase $\sigma$ for $\gamma = 64^\circ$ and $\delta 
= 0$. 
Solid, dotted, dashed and dotdashed lines are for 
$K^-\pi^+$, $\bar K^0 \pi^-$, $K^-\pi^0$ and $\phi K$ respectively.
} 
\end{figure}

For physics beyond SM such as $c_8 = 2 e^{i\sigma}$, 
there are too many parameters and one needs a strategy.
We set $N = 3$ and try to fit observed $Br$'s with
the phase $\sigma$, then find the preferred $a_{CP}$.
Since the $c_8$ term now dominates,
one is less sensitive to the FSI phase $\delta$.
In fitting $Br$'s, we find that 
{\it destructive interference is necessary} which can be understood 
from the inclusive results of Fig. 1.
This means that {\it large $a_{CP}$s are preferred!}
We plot in Fig. 2(e) and (f) the $Br$ and $a_{CP}$
vs. the new physics phase $\sigma$,
for $\gamma = 64^\circ$ and $\delta = 0$.
The $K^-\pi^+$ and $\bar K^0 \pi^-$ modes are very close in rate
for $\sigma \sim 45^\circ - 180^\circ$,
but the $K^-\pi^0$ mode remains a factor of 2 smaller.
However,
the $a_{CP}$ can now reach 50\% for $K^-\pi^+$/$\bar K^0 \pi^-$
and 40\% for $K^-\pi^0$!
These are {\it truly large asymmetries} and would be easily observed,
perhaps even before B Factories turn on (i.e. with CLEO II.V data!).
They are in strong contrast to the SM case with FSI phase $\delta$,
Fig. 2(d), and can be distinguished.

Genuine pure penguin processes arising from $b\to s\bar ss$
give cleaner probes of new physics $CP$ violation effects
since they are insensitive to FSI phase.
The amplitude for $B^-\to \phi K^-$ decay is
\begin{eqnarray}
A(B\rightarrow \phi K)&\simeq&
-i {G_F\over \sqrt{2}} f_\phi m_\phi
                       2 p_B\cdot \varepsilon_\phi F_1(m_\phi^2)\, 
\left\{ v_j \left(c_{3}^j+{c_4^j\over N} + c_5^j\right) \right|_{q_2^2}
\nonumber\\
 && \left. +~v_j \left({c_3^j\over N} +c_4^j + {c_6^j\over 
N}\right)\right|_{q_1^2}
        + \left.
 v_t {\alpha_s\over 4\pi} {m_b^2 \over q_1^2} c_8 \tilde S_{\phi 
K} \right\}.
\label{phiK}
\end{eqnarray}
The relevant $q^2$ is determined by kinematics:
$q_1^2 = m_b^2/2$ as before,
but for amplitudes without Fierzing $q_2^2 = m_\phi^2$.
We have dropped color octet contributions
and have checked that they are indeed small.
Since the amplitude is pure penguin, $c_8$ should have no absorptive part.
As shown in Fig. 2(a) and (b),
the SM rate of $\sim 1\times 10^{-5}$ is above the CLEO bound 
of $0.5\times 10^{-5}$ while $a_{CP}$ is uninterestingly small.
If we allow for new physics enhanced $c_8 = 2 e^{i\sigma}$,
one again needs destructive interference to match observed rate.
The results are plotted in Fig. 2(e) and (f) vs. $\sigma$.
The rate is lower than the $\bar K^0 \pi^-$/$K^-\pi^+$ modes
because it is not sensitive to $1/m_s$ and we have used a low
$m_s$ value to boost up $B\to K\pi$.
The $a_{CP}$ could now reach almost 60\%,
thanks to the destructive interference preferred by 
fitting the CLEO limit on rate.
We note that the SM asymmetry for $B\to \phi K$ should be of order 1\%.

One can now construct an attractive picture.
We have noted that recent studies cannot explain
the low  $B\to \phi K$ upper limit within SM.
If $c_8$ is enhanced by new physics and interferes destructively with SM,
$B\to \phi K$ can be brought down to below $5\times 10^{-6}$.
The experimentally observed $\bar K^0 \pi^- \simeq K^-\pi^+$
follows from $c_8$ dominance,
and their rate $\sim 1.4\times 10^{-5}$ which is 
2--3 times larger than the $\phi K$ mode suggests 
a low $m_s$ value and slight tunings of BSW form factors.
Around $\sigma \sim 145^\circ$,
the rates are largely accounted for,
but $a_{CP}$ for $\phi K$, $K^-\pi^+/K^-\pi^0$ and $\bar K^0 \pi^-$
could be {\it enhanced to the dramatic values} of
55\%, 45\% and 35\%, respectively,
and {\it all of the same sign}.
This is certainly distinct from the sign correlations of SM with FSI.

On the down side, within the scenario of strong penguin dominance, 
which includes the case of enhanced $c_8$,
the $B^-\to K^-\pi^0$ rate is always about a factor of two smaller
than the $K^-\pi^+$ mode, and we are unable to accommodate
recent CLEO findings.~\cite{cleo1,DHHP}
We are also barely able to accommodate $B\to \omega K$.
Within SM one needs $1/N_{\sst\rm eff} \sim 1$ to be able to
account for the large $B\to \omega K \simeq 1.5\times 10^{-5}$ value,
while for $1/N_{\sst\rm eff} \simeq 0$ one can account for only half.
Adding new physics induced  $c_8 = 2 e^{i\sigma}$ effect,
we are able to account for $Br$ for both large and small $N_{\sst\rm eff}$,
but not for $N = 3$. 
However, $a_{CP}$ is never more than a few \%
and hence not very interesting.
Since the $ \omega K$ mode has a single isospin amplitude,
it is insensitive to FSI rescattering phases.

\section{CP Violation in $b\to s\gamma$ Decays}

We have emphasized that the $b\to s$ modes that are 
already observed are the best places for CP search.
Clearly, the $B\to K^*\gamma$ and $b\to s\gamma$ modes were
the first ever observed penguins in $B$ decay,
and they should provide a good window.
We note that the observed recoil $m_{X_s}$ spectrum
for $B \to \gamma + X_s$ is basically orthogonal to
that for $B\to \eta^\prime +X_s$,
and is clearly dominated by $K$ resonances.
However, besides the $B\to K^*\gamma$ mode,
the higher resonance contributions to the inclusive spectrum 
has not yet been disentangled.

It is important to realize that,
although in our previous discussions of enhanced $b\to sg$, 
one must reckon with the $b\to s\gamma$ constraint,
{\it the converse is not true}.
One can have interesting impact on $b\to s\gamma$ 
without affecting $b\to sg$ by much.
Our example of $\tilde s_{L,R} - \tilde b_{L,R}$ mixings
can generate a lot of effects.
The $c_7 O_7$ operator structure of SM
can become $c_7 O_7 + c_7^\prime O_7^\prime$,
where $O_7^\prime$ has opposite chirality to $O_7$
(and likewise for the gluonic $O_8$).
This leads to much enrichment of the physics compared to SM:~\cite{CHH}

\begin{itemize}
\item Direct CP violation in $b\to s\gamma$ and $B\to K^*\gamma$

Since SM accounts for the observed $b\to s\gamma$ rate already,
one has the constraint
$\vert c_7^{\sst\rm SM}\vert^2 \sim
\vert c_7^{\sst\rm SM} + c_7^{\sst\rm New}\vert^2 + \vert c_7^{\prime}\vert^2$,
hence one prefers $c_7^{\prime}$ to be small.
We find that $a_{CP}$ larger than 10\% is possible in certain parameter space,
especially when the new physics effect has opposite sign w.r.t. SM.
\item Mixing dependent CP violation

This requires interference between $O_7$ and $O_7^\prime$,
the two different chiralities.
For $B^0\to M^0\gamma$, where $M^0$ is a 
CP eigenstate with eigenvalue $\xi$, one obtains a 
mixing dependent asymmetry~\cite{AGS}
\[
A_{\rm mix} =2\xi \,\frac{|c_{7}c_{7}^{\prime}|}
{|c_{7}|^{2}+|c_{7}^{\prime}|^{2}}\,\sin [\phi _{B}-\phi -\phi ^{\prime}],
\]
where $\phi ^{(\prime)}$ are the weak phases of $c_{7}^{(\prime)}$.
Note that $\phi ^{(\prime)}$ could vanish and one could still have 
CP violation through $\phi_B$ from $B$-mixing within SM.
We find that the coefficient to the phase factor can reach 80\%
in special regions of parameter space of our model. 
Unfortunately,
$B^0\to K^{*0}\gamma \to K_S\pi^0\gamma$ does not give a vertex,
and one would need to turn to either $B_s\to \phi\gamma$,
or $B_d\to K_1^0\gamma$, $K_2^{*0}\gamma$,
none of which are yet observed.
\item Chirality probe: $\Lambda_b \to \Lambda\gamma$

How does one know that both SM-like
$b_{\sst R}\to s_{\sst L}\gamma_{\sst {\rm ``}L{\rm "}}$ 
and new physics induced
$b_{\sst L}\to s_{\sst R}\gamma_{\sst {\rm ``}R{\rm "}}$ transitions occur?
The best way, independent of CP violation
(but direct $a_{CP}$ in rates is of course possible),
is to search for $\Lambda_b \to \Lambda\gamma$ decay,
since $\Lambda \to p\pi$ decay is self-analyzing.
One has~\cite{MR}
\[
{d\Gamma \over d\cos \theta} \propto 1+
{|c_{7}|^{2}-|c_{7}^{\prime }|^{2}\over |c_{7}|^{2}+|c_{7}^{\prime }|^{2}}
\cos \theta,
\]
where $\theta$ is the angle between 
the direction of $\vec{p}_\Lambda$ 
in $\Lambda _{b}$ frame and 
the direction of the $\Lambda$ polarization in its rest frame.
The coefficient of $\cos\theta$ is clearly equal to 1 in SM,
but could be different in Nature.
We find that even $-1$ is possible!
When and where will $\Lambda_b\to \Lambda\gamma$ decay
be measured?
\end{itemize}

\section{Discussion and Conclusion}

We must recall that $B$ physics has had its share of surprises.
The long $b$ lifetime was discovered without much theory encouragement.
$B_d$ mixing was in fact discovered with theory ``discouragement".
More recently,
the $\eta^\prime K$ and $\omega K$ modes turn out to be 
much larger than theory expectations,
while the huge inclusive fast $\eta^\prime + X_s$
simply came out of the blue without theory warnings.
We therefore must be on guard for CP violation.

In the narrow sense, we have discussed a large 
$\tilde s -\tilde b$ squark mixing model that could generate 
large color dipole $bsg$ coupling which carries an
unconstrained new CP phase, 
and lead to large impact on CP violating asymmetries:
in $\eta^\prime + X_s$,
charmless 2-body modes such as $K\pi$ and $\phi K$,
$b\to s\gamma$,
even~\cite{HH} in $J/\psi K_S\pi^0$ modes.
In the broad sense, we have
illustrated that {\it large CP asymmetries
may just pop up everywhere} as B Factories turn on!

Let's search for CP violation in already observed modes,
{\it assuming they are large}!

\section*{Acknowledgments}
I thank
Chun-Khiang Chua, Xiao-Gang He, Ben Tseng and
Kwei-Chou Yang for collaborative work,
and R.O.C. National Science Council for support.

\end{document}